\documentclass[8pt]{article}

\usepackage{amsmath,amssymb,amsthm,graphicx,latexsym}

\newcommand{\ed}{\end{document}}
\newcommand{\beq}{\begin{equation}}
\newcommand{\eeq}{\end{equation}}
\newcommand{\beqa}{\begin{eqnarray}}
\newcommand{\eeqa}{\end{eqnarray}}
\newcommand{\bc}{\begin{center}}
\newcommand{\ec}{\end{center}}
\newcommand{\vs}{\vspace}
\newcommand{\ba}{\begin{array}}
\newcommand{\ea}{\end{array}}

\begin{document}
\begin{center}
\large{\textbf{\textbf{Electrodynamics of a Generalized  Charged
Particle in Doubly Special Relativity Framework}}}\\
\end{center}
\bc
Souvik Pramanik \footnote{E-mail: $souvick.in@gmail.com$},
Subir Ghosh \footnote{E-mail: $subir_{-}ghosh2@rediffmail.com$}\\
Physics and Applied Mathematics Unit, Indian Statistical
Institute\\
203 B. T. Road, Kolkata 700108, India \\
and\\
Probir Pal  \footnote{E-mail: $probirkumarpal@rediffmail.com$}\\
S. N. Bose National Centre for Basic Sciences,\\
 Block-JD, Sector-III, Salt Lake, Kolkata 700 098, India
\ec
\vs{0.5cm}
\begin{center}
{\textbf{Abstract:}}
\end{center}
\emph{In the present paper, dynamics of generalized charged particles are studied in the presence of external electromagnetic interactions. This particular extension of the free relativistic particle model lives in Non-Commutative $\kappa$-Minkowski space-time, compatible with Doubly Special Relativity, that is motivated to describe Quantum Gravity effects.
Furthermore we have also considered the electromagnetic field to be dynamical and have derived the modified forms of Lienard-Wiechert like potentials for these extended charged particle models.
In all the above cases we exploit the new and extended form of $\kappa$-Minkowski algebra where electromagnetic effects are incorporated in the lowest order, in the Dirac framework of Hamiltonian constraint analysis.}
\vskip .3cm
\section{Introduction}
Non-Commutative (NC) geometry effects in physics have developed into an important area of research in recent times. Different structural forms of NC geometry are motivated to describe different effects. In the present work we concentrate on a particular NC phase space that obeys $\kappa$-Minkowski NC algebra \cite{dsr2}{\footnote{The NC geometry was introduced in physics very early by \cite{sn} and resurrected to its present interest by the work of \cite{sw}.}}. The $\kappa$-Minkowski NC algebra is compatible with Doubly Special Relativity (DSR) framework, introduced  by Amelino-Camelia \cite{am}. (For reviews see for example \cite{am1}). DSR is motivated to describe Quantum Gravity effects. (We will briefly return to this point towards the end of the paper.) Out of several alternative bases of the algebra \cite{kow}, that are inter-related through non-linear transformations, we will focus on the Magueijo-Smolin (MS) base \cite{mag}. In general, extended relativistic principles in DSR requires a modification in the Einstein energy-momentum relation $P_\mu P^\mu=m_0^2$ for a free particle of rest mass $m_0$ and momentum $P_\mu$. Quite naturally the canonical Lagrangian $L=m_0{\sqrt{((dX_\mu)/(d\tau ))^2}}$ also needs to be drastically altered to account for the DSR particle dispersion  relation and $\kappa$-Minkowski NC phase space. Two of the present authors \cite{pal}, along with many other workers \cite{others}, have provided explicit models for DSR particles that is consistent with the $\kappa$-Minkowski NC algebra and DSR dispersion relation. Many novel features of the DSR particle have come into light from the analysis of these models.

It needs to be emphasized that mostly these works deal with the free DSR particle but many interesting properties as well as drawbacks of the model (if there are any) are not encountered unless the particle is subjected to non-trivial interactions. In the present paper we have precisely done that. We have considered a DSR particle interacting with a $U(1)$ gauge theory. We have mostly concentrated on treating the gauge field as external but we have also commented on the DSR effects on the Lienard-Wichert type potentials arising from the charged DSR particle.

The paper is organized as follows: in Section 2 we introduce the $\kappa$-Minkowski NC algebra in MS base and the interacting DSR particle - $U(1)$ gauge theory. In Section 3 we derive the extended $\kappa$-Minkowski algebra in the presence of  $U(1)$ gauge interaction and obtain the equations of motion. Section 4 deals with explicit solutions for particle trajectories in a perturbative framework, with only electric field, that provide valuable insights of the DSR effects on a charged particle motion. Section 5 considers the MS particle dynamics with both electric and magnetic fields present. In Section 6 we comment on the electromagnetic potentials induced by the charged DSR particle. The paper concludes in Section 7 with discussions and future outlook.
\section{Free $\kappa$-Minkowski algebra, free MS particle kinematics and
modeling interacting MS particle in external $U(1)$ field}
The $\kappa$-Minkowski NC phase space algebra is defined as,
\begin{equation}
\{x^i,x^0\}=\frac{x^i}{\kappa} ~;~~
\{x^i,x^j\}=0~;~~\{x^i,p^j\}=-g^{ij}~;~\{p^\mu,p^\nu\}=0.
\label{ko}
\end{equation}
Our metric is $diag~g^{00}=-g^{ii}=1$ and $\kappa $ is the  NC parameter. Rest of the phase space algebra is given below,
\begin{equation}
 \{x^0,p^i\}=p^i/\kappa ~;~\{x^i,p^0\}=0~;~\{x^0,p^0\}=-1+p^0/\kappa.
\label{2}
\end{equation}
The above is rewritten in a covariant form,
$$
\{x_\mu ,x_\nu \}=\frac{1}{\kappa}(x_\mu \eta_{\nu}-x_\nu
\eta_{\mu }),$$
\begin{equation}
\{x_{\mu},p_{\nu}\}=-g_{\mu\nu}+\frac{1}{\kappa}\eta_{\mu}p_{\nu},~~\{p_{\mu},p_
{\nu}\}=0,
 \label{03}
\end{equation}
where $\eta _0=1,\eta _i=0$. This algebra appeared in \cite{granik1}. Detailed studies of similar types of algebra are provided in \cite{kow}.  This algebra has emerged before in an earlier work of one of us \cite{sg1} where it was embedded in a more general algebra. For $\kappa \rightarrow \infty $ one recovers the normal canonical phase space. Using $\kappa$-Minkowski NC algebra (\ref{03}), it is straightforward to check that the Lorentz generators $J_{\mu\nu }=x_\mu p_\nu -x_\nu p_\mu $ satisfy the canonical Lorentz algebra $J^{\mu\nu }$,
\begin{equation}
\{J^{\mu\nu },J^{\alpha\beta }\}=g^{\mu\beta }J^{\nu\alpha
}+g^{\mu\alpha }J^{\beta \nu}+g^{\nu\beta }J^{\alpha\mu
}+g^{\nu\alpha }J^{\mu\beta }. \label{51}
\end{equation}
However the particle dispersion relation that remains invariant under (\ref{03}) with respect to $J^{\mu\nu}$ treated as
rotation generators in the NC framework is given below:
\begin{equation}
p^2=m^2\left[1-\frac{(\eta p)}{\kappa} \right]^2 = m^2\left[1-\frac{E}{\kappa}
\right]^2.  \label{m}
\end{equation}
We use the shorthand notation $(ab)\equiv a^\mu b_\mu$. This is the well known MS dispersion relation that reduces to the Special Theory relation for $\kappa \rightarrow \infty $ that is the energy upper bound (induced by DSR) is pushed back to infinity.

The fact that it is possible to construct explicitly generators that satisfy an undeformed canonical Lorentz algebra means that DSR corrections do not affect the Lorentz group multiplet structure \cite{mag}. However it is important to remember $x_\mu$ and $p_\mu $ do not transform canonically under infinitesimal Lorentz transformations $(\{x_\mu, J^{\nu\lambda }\},~\{p_\mu, J^{\nu\lambda}\})$ and finite Lorentz transformations are affected by DSR corrections as well (for details see \cite{pal}). Hence the model is invariant with respect to generalized Lorentz transformations. One can see this explicitly from checking that $\{p^2/[1-\frac{(\eta p)}{\kappa}]^2,J^{\nu\lambda }\}=0$ thereby ushering in the DSR dispersion relation (\ref{m}).

In order to derive a Lagrangian for the DSR particle {\it{ab initio}} it is quite difficult to guess how to proceed since it has to be compatible with the complicated $\kappa$-Minkowski symplectic structure (\ref{03}). In this perspective, it is very convenient to exploit the Darboux-like canonical variables. The idea is to construct the Lagrangian in terms of canonical variables which are in fact the Darboux map of the physical non-canonical (or NC) degrees of freedom. The next step is to reexpress the the Lagrangian in term of the physical NC variables via the inverse Darboux map. We have successfully exploited this technique in a previous work \cite{pal}. In the present case, with the canonical phase space
\begin{equation}
\{X_\mu ,P_\nu \}=-g_{\mu\nu}~,~~\{X_\mu ,X_\nu \}=\{P_\mu ,P_\nu\}=0, \label{c2}
\end{equation}
the inverse Darboux map is
\begin{equation}
X_\mu \equiv x_\mu \left(1-\frac{(\eta p)}{\kappa}\right)=x_\mu
\left(1-\frac{E}{\kappa}\right);~~P_\mu \equiv \frac{p_\mu}{\left(1-\frac{(\eta
p)}{\kappa}\right)}=\frac{p_\mu}{\left(1-\frac{E}{\kappa}\right)}, \label{c1}
\end{equation}
whereas the Darboux map is,
\begin{equation}
x_\mu = X_\mu \left(1+\frac{(\eta P)}{\kappa}\right)=X_\mu
\left(1+\frac{P_0}{\kappa}\right);~~p_\mu = \frac{P_\mu}{\left(1+\frac{(\eta
P)}{\kappa}\right)}=\frac{P_\mu}{\left(1+\frac{P_0}{\kappa}\right)}. \label{c3}
\end{equation}
We consider the first order form of a particle interacting minimally with $U(1)$ gauge field,
\begin{eqnarray}
L &=& P^\mu(\tau ) \dot{X}_\mu(\tau )-\lambda(P^2(\tau )-m_0^2)+\int d^4y~e~\delta (y-X(\tau ))\dot{X}_\mu A^\mu (y)\nonumber\\
&=& P^\mu(\tau ) \dot{X}_\mu(\tau )-\lambda(P^2(\tau )-m^2)+ ~e~\dot{X}_\mu(\tau)A^\mu X(\tau)\nonumber\\
&=& (p\dot{x})-\frac{(px)(\eta\dot{p})}{\kappa(1-\frac{(\eta p)}{\kappa})}-\frac{\lambda}{2}\left(p^2-m^2\left(1-\frac{(\eta p)}{\kappa}\right)^2\right)-eA^\mu(X)\left(\dot{x}_\mu\left(1-\frac{(\eta p)}{\kappa}\right)-x_\mu\frac{(\eta\dot{p})}{\kappa}\right).\label{lagr}
\end{eqnarray}
In the last equality we have used the inverse Darboux map (\ref{c1}) and the notation $a^\mu b_\mu \equiv (ab)$. As has been shown before \cite{pal}, for the free theory $e=0$, Dirac constraint analysis \cite{dirac} yields the NC $\kappa$-Minkowski symplectic structure. Since the transformation (\ref{c3}) is non-canonical, the dynamics of the particle in two representations $(X,P)$ and $(x,p)$ are inequivalent.
\section{Extended $\kappa$-Minkowski algebra with gauge interactions and MS particle dynamics}
Our first task is to compute the symplectic structure modified by the electromagnetic interaction. We utilize the Dirac formalism \cite{dirac} for Hamiltonian analysis of constraint system. A similar analysis for the free theory $(e=0)$, was done in \cite{pal}. From (\ref{lagr}) the constraints are,
\beq \eta_\mu^1\equiv \Pi^x_\mu-p_\mu-e\psi A_\mu \approx 0,\eeq
\beq \eta_\mu^2\equiv \Pi^p_\mu+\frac{(px)\eta_\mu}{\kappa\psi}+\frac{e(x
A)\eta_\mu}{\kappa}\approx 0,\eeq
where $\psi \equiv 1-\frac{(\eta p)}{\kappa}$ and the conjugate momenta are
\beq \Pi^x_\mu=\frac{\partial L}{\partial \dot{x}_\mu}=p_\mu+e\psi A_\mu~;~~
\Pi^p_\mu=\frac{\partial L}{\partial
\dot{p}_\mu}=-\frac{(px)\eta_\mu}{\kappa\psi}-\frac{e(x A)\eta_\mu}{\kappa}.\eeq
The constraints $\eta^1_\mu, \eta^2_\mu$ are non-commuting since
\beq \{\eta_\mu^1,\eta_\nu^1\}=e\psi^2\left(\frac{\partial A_\mu}{\partial
X_\nu}-\frac{\partial A_\nu}{\partial X_\mu}\right)=e\psi^2 F_{\mu\nu},\eeq
\beq
\{\eta_\mu^1,\eta_\nu^2\}=g_{\mu\nu}+\frac{p_\mu\eta_\nu}{\kappa\psi}+\frac{
e\psi}{\kappa}F_{\mu\lambda}x^\lambda\eta_\nu \eeq
\beq \{\eta_\mu^2,\eta_\nu^2\}=\frac{(x_\mu\eta_\nu-x_\nu\eta_\mu)}{\kappa\psi}.
\eeq
Hence the constraints are Second Class in Dirac sense and we wish to compute the
Dirac bracket \cite{dirac}, as defined below,
\beq \{A,B\}^*=\{A,B\}-\{A,\eta^i\}\{\eta^i,\eta^j\}^{-1}\{\eta^j,B\},\eeq
so that the constraints can be strongly put to zero. Expressing the constraint
matrix as
\begin{equation}
\ba{lcl}
  \{\eta_\mu^i,\eta_\nu^j\}=\left[ {\begin{array}{cc}
 0 & g_{\mu\nu}+\frac{p_\mu\eta_\nu}{\kappa\psi}  \\
 -g_{\mu\nu}-\frac{p_\nu\eta_\mu}{\kappa\psi} &
\frac{(x_\mu\eta_\nu-x_\nu\eta_\mu)}{\kappa\psi}  \\
 \end{array} } \right]
 +e\left[ {\begin{array}{cc}
 \psi^2F_{\mu\nu} & \frac{\psi}{\kappa}F_{\mu\lambda}x^\lambda\eta_\nu  \\
 -\frac{\psi}{\kappa}F_{\nu\lambda}x^\lambda\eta_\mu  & 0  \\
 \end{array} } \right]$$$$
 \equiv A+eB
 \ea
 \end{equation}
and using the inverse $$ (A+eB)^{-1}=A^{-1}-e(A^{-1}B A^{-1}) +O(e^2)$$ the required inverse of the constraint matrix to $O(e)$ is computed below:
\begin{equation}
 \{\eta_\nu^i,\eta_\lambda^j\}^{-1}=\left[ {\begin{array}{cc}
 \frac{1}{\kappa}(x_\nu\eta_\lambda-x_\lambda\eta_\nu) &
-g_{\nu\lambda}+\frac{1}{\kappa}\eta_\nu
p_\lambda+\frac{e\psi^2}{\kappa}x_\nu\eta^\alpha F_{\alpha\lambda} \\
 g_{\nu\lambda}-\frac{1}{\kappa}\eta_\lambda
p_\nu-\frac{e\psi^2}{\kappa}x_\lambda\eta^\alpha F_{\alpha\nu} &
e\psi^2[F_{\nu\lambda}+\frac{\eta^\alpha}{\kappa}(p_\nu
F_{\alpha\lambda}-p_\lambda F_{\alpha\nu})] \\
 \end{array} }\right].
\end{equation}
Therefore to $O(e)$ the Dirac Brackets are
\begin{eqnarray}
\{x_\alpha,x_\beta\}^*=\{x_\alpha,x_\beta\}-\{x_\alpha,\eta_\nu^1\}\{\eta_\nu^1,\eta_\lambda^1\}^{-1}\{\eta_\lambda^1,x_\beta\}
&=&\frac{1}{\kappa}(x_\alpha\eta_\beta-x_\beta\eta_\alpha)\label{d1}\\
\{x_\alpha,p_\beta\}^*=\{x_\alpha,p_\beta\}-\{x_\alpha,\eta_\nu^1\}\{\eta_\nu^1,\eta_\lambda^2\}^{-1}\{\eta_\lambda^2,p_\beta\}
&=&\left(-g_{\alpha\beta}+\frac{1}{\kappa}\eta_\alpha p_\beta\right)+\frac{e\psi^2}{\kappa}x_\alpha\eta^\nu F_{\nu\beta} \label{d2}\\
\{p_\alpha,p_\beta\}^*=\{p_\alpha,p_\beta\}-\{p_\alpha,\eta_\nu^2\}\{\eta_\nu^2,\eta_\lambda^2\}^{-1}\{\eta_\lambda^2,p_\beta\}
&=& e\psi^2\left(F_{\alpha \beta}+\frac{\eta^\nu}{\kappa}(p_\alpha F_{\nu\beta}-p_\beta F_{\nu\alpha})\right).\label{all}
\end{eqnarray}
For $e=0$ the above reduces to the free $\kappa $-Minkowski algebra \cite{kow}. It is worth pointing out that $\{x_\alpha,x_\beta \}$ remains unchanged and the modifications involve the gauge invariant field tensor $F_{\mu\nu}(X)$.

To obtain the equations of motion we write the Hamiltonian first in terms of canonical variables and subsequently in terms of NC variables:
\beq
H=\frac{P^2}{m_0}-\sqrt{P^2}=\frac{p^2}{m_0\psi^2}-\frac{\sqrt{p^2}}{\psi}\label {hh}\eeq
As $P^2=m_0^2\Rightarrow p^2=m_0^2\psi^2$. The bracket (\ref{all}) is non-vanishing (hence non-canonical) even when DSR effects are absent, $\kappa^{-1}=0$, $$\{p_\alpha,p_\beta\}^*=eF_{\alpha \beta},\{x_\alpha,p_\beta\}^*= -g_{\alpha\beta},~\{x_\alpha,p_\beta\}^*=0 .$$ This algebra, together with a free Hamiltonian, will reproduce the normal (non-DSR) electrodynamics of a charged particle. The DSR electrodynamics will be recovered if we use the $\kappa$-dependent algebra (\ref{d1},\ref{d2},\ref{all}). The other alternative (for non-DSR case) is to use purely canonical Poisson brackets,
$$\{p_\alpha,p_\beta\}=0,~\{x_\alpha,p_\beta\}=-g_{\alpha\beta},~\{x_\alpha,p_\beta\}=0$$ and an interacting Hamiltonian $\sim (p_\mu -eA_{\mu})^2$. In the present DSR-modified dynamics we employ the former procedure.

An important feature of the DSR effect is that it is $U(1)$ gauge invariant. This is clearly seen from the $*$-algebra which depends on the gauge invariant structure $F_{\mu\nu}$.

Note that this $H$ should not be identified as the energy but it is easy to check that in the canonical sector it clearly reproduces the correct Hamiltonian equations of motion,
\beq \frac{d x^\alpha}{d\tau}\equiv\dot{x}^\alpha=\{x^\alpha,H\}^*~,~~~
\frac{d p^\alpha}{d\tau}\equiv\dot{p}^\alpha=\{p^\alpha,H\}^*.\eeq

We should mention that there is an approximation involved in deriving $H$ in (\ref{hh}). Strictly speaking we should have used the Darboux map pertaining to the $U(1)$ modified algebra (\ref{d1}-\ref{all}) whereas, mainly for simplicity we have used the Darboux map (\ref{c1},\ref{c3}) for the free NC algebra (\ref{03}) as derived in \cite{pal}. Although $O(e)$ correction to the Darboux map is straightforward to compute we have not considered this here. But as we demonstrate in rest of the paper, even in this relatively simplified situation, the $\kappa$-corrections generate strikingly qualitative differences.

Therefore the equations of motion  to $0(e)$ follow:
\begin{eqnarray}
\dot{x}^\alpha &=& (2-\psi)\frac{1}{m_0}\left[-\frac{1}{\psi ^2}p^\alpha+\frac{e}{\kappa}(\eta F p)x^\alpha\right],\label{eqm1}\\
\dot{p}^\alpha &=& (2-\psi)\frac{e}{m_0}\left[F^{\alpha\beta}p_\beta+\frac{1}{\kappa}(\eta F p)p^\alpha +2m_0^2F^{\alpha\beta}\eta_\beta\right].\label{eqm2}
\end{eqnarray}

Let us write down the modified Lorentz force equation to order $0(e)$. We differentiate (\ref{eqm1}) with respect to $\tau$ to obtain,
\beq \ddot{x}^\alpha = (2-\psi)\frac{1}{m_0}\left[-\frac{1}{\psi ^2}\dot p^\alpha+\frac{e}{\kappa}(\eta F p)\dot
x^\alpha +\frac{2\dot \psi}{m_0\psi ^3}p^\alpha\right]+\frac{\dot \psi}{m_0\psi^2}p^\alpha.\label{e1}\eeq

Replacing all the $p$ dependent terms in the R.H.S of (26) by $\dot{x}$ (using (24)) and keeping terms only upto O(e) we finally obtain
\beq m_0 \ddot{x}^\alpha =eF^{\alpha\beta}p_\beta+\frac{e m_0}{\kappa}
\left[-\frac{2(2-\psi)^2}{\psi^2}F^{\alpha\beta}\eta_\beta+\frac{(4-\psi)\psi^4}{(2-\psi)}(\eta F\dot x)
\left(\frac{\psi^2m_0}{(2-\psi)}(\eta \dot x)-1\right)\dot x^\alpha-\psi^2(\eta F\dot x)\dot x^\alpha \right].
\label{e2}\eeq

Clearly the MS Lorentz force law is quite involved with non-linear terms coming in through the $\kappa$-correction and for $\kappa \rightarrow\infty$ the conventional Lorentz law is smoothly recovered.
\section{Trajectories of charged MS particles in external electric field}
We will try to solve perturbatively the first order $\tau$-derivative force equation (\ref{eqm2}) for some specific external field configuration. Due to the nature of the basic (MS form of $\kappa$-Minkowski algebra in (\ref{03})) NC structure  considered, it is easy to see that the $\kappa$-corrections always involve the electric field in the form $(\eta F)_\alpha$, as is clear from (\ref{d1}-\ref{all}). Hence for situations with external purely magnetic fields the dynamics do not change very much in a qualitative way. Hence for the present case, we consider only external constant electric field. Again this analysis is true for combined electric and magnetic fields but with the condition $\mid \vec E\mid >\mid \vec B\mid >$ such that one can Lorentz transform to another frame with  electric field only (see for example \cite{jac}). Without loss of generality, let us consider the electric field  to be along the direction of $x_1$-axis, $E_1 = E,~E_2=E_3=0$. Breaking up the set of equations in (\ref{eqm2}), we find
\begin{eqnarray}
\frac{d p_0}{d\tau}=\frac{e E}{m}p_1+\frac{2 e E}{m\kappa}p_0p_1,~
\frac{d p_1}{d\tau}=\frac{e E}{m}p_0+\frac{e E}{m\kappa}(p_0^2+2m^2-p_1^2),~
\frac{d p_2}{d\tau}=-\frac{e E}{m\kappa}p_1p_2,~
\frac{d p_3}{d\tau}=-\frac{e E}{m\kappa}p_1p_3. \label{p3}
\end{eqnarray}

We immediately notice a qualitative difference: For the normal particle the force is impressed only along the direction of the external electric field and the energy changes accordingly. As for example in the present case the electric field is along $x_1$ and for the normal particle the momentum will change only $x_1$ direction. However, for the MS particle we observe that not only are the rates of changes energy and $x_1$-momentum  modified, but more interestingly momenta along $x_2$ and $x_3$ directions are also affected. Note that the mass $m$ will actually depend on the initial velocity as we specify below. We further differentiate the above set of equations and keeping terms upto the first order term of $\frac{1}{\kappa}$, a little rearrangement leads to:
\begin{eqnarray}
\frac{d^2p_0}{d\tau^2} &=& \epsilon^2\left[p_0+\frac{1}{\kappa}\left(3p_0^2+p_1^2+2m^2\right)\right]\label{d2p0}\\
\frac{d^2p_1}{d\tau^2} &=& \epsilon^2\left[p_1+\frac{2}{\kappa}p_0p_1\right]\label{d2p1}\\
\frac{d^2 p_2}{d\tau^2} &=& -\frac{\epsilon^2}{\kappa}p_0p_2\label{d2p2}\\
\frac{d^2 p_3}{d\tau^2} &=& -\frac{\epsilon^2}{\kappa}p_0p_3\label{d2p3}
\end{eqnarray}
where $\frac{e E}{m}=\epsilon$. This is the system of non-linear ordinary differential equations we plan to solve
to the leading order in $1/\kappa $. Let the initial conditions for the momenta at $\tau=0$ be  $p_0(0)=m,~p_2(0)=mv_0,~ p_1(0)=p_3(0)=0$ so that $m=\frac{m_0}{\sqrt{1-V_0^2}}, v_0=\frac{V_0}{\sqrt{1-V_0^2}}$ where $V_0=\frac{dx}{dt}|_{t=0},~
v_0=\frac{dx}{d\tau}|_{\tau=0}$. We provide some details of solving perturbatively the non-linear set of equations (\ref{d2p0}-\ref{d2p3}) in the Appendix. To $O(1/\kappa )$ the solutions are,
\begin{eqnarray}
p_0 &=& m~Cosh(\epsilon\tau)+\frac{14 m^2}{6\kappa}~Cosh(\epsilon\tau)+\frac{2m^2}{3\kappa}~Cosh(2\epsilon\tau)
-\frac{3 m^2}{\kappa},\nonumber\\
p_1 &=& m~Sinh(\epsilon\tau)+\frac{14 m^2}{6\kappa}~Sinh(\epsilon\tau)+\frac{m^2}{3\kappa}~Sinh(2\epsilon\tau),\nonumber\\
p_2 &=& m v_0\left[1+\frac{m}{\kappa}\left(1-Cosh(\epsilon\tau)\right)\right],~p_3=0. \label{p3final}
\end{eqnarray}

Since $p_0=m\frac{d x_0}{d\tau}$ and $p_i=m\frac{d x_i}{d\tau}$, integrating $p_\mu$ from the above set of equations with respect to $\tau$ gives $x_\mu$ as
\begin{eqnarray}
\epsilon x_0 &=& \left(1+\frac{7 m}{3\kappa}\right)~Sinh(\epsilon\tau)
+\frac{m}{3\kappa}~Sinh (2\epsilon\tau)-\frac{3m \epsilon\tau}{\kappa},\label{equx0}\\
\epsilon x_1 &=& \left(1+\frac{7 m}{3\kappa}\right)~Cosh(\epsilon\tau)
+\frac{m}{6\kappa}~Cosh(2\epsilon\tau)-\left(1+\frac{5m}{2\kappa}\right),\label{equx1}\\
\epsilon x_2 &=& v_0\left[\epsilon\tau\left(1+\frac{m}{\kappa}\right)-\frac{m}{\kappa}~Sinh(\epsilon\tau)\right],
\label{equx2}\\
x_3 &=& 0. \label{equx3}
\end{eqnarray}

Hence we have found the paths of the MS particle in proper time. Finally we can eliminate $\tau$ from (\ref{equx1}) and (\ref{equx2}) to construct the trajectories of the particle as,
\beq \left\{x_1+\frac{m}{e E}\left(1+\frac{5m}{2\kappa}\right)\right\}^2-\frac{x_2^2}{v_0^2}\left(1+\frac{11m}{2\kappa}\right)
-\frac{x_2^4}{v_0^4}\frac{e^2E^2}{3m\kappa}
=\frac{m^2}{e^2E^2}\left(1+\frac{5m}{\kappa}\right). \label{equx1x2}\eeq

The $\kappa$-corrected  hyperbolic path is recovered provided we neglect the $x_2^4$ term as it is of $O(e^2/\kappa )$. This leads to a modified hyperbola:
\beq \left\{\frac{e E}{m}x_1+\left(1+\frac{5m}{2\kappa}\right)\right\}^2-\frac{e^2E^2}{m^2v_0^2}\left(1+\frac{11m}{2\kappa}\right)x_2^2
=\left(1+\frac{5m}{\kappa}\right) \label{equhyper} \eeq

The eccentricity $\Sigma$ is reduced by the $\kappa$-correction,
\beq
\Sigma=\sqrt{1+v_0^2-\frac{11m v_0^2}{2\kappa}}.\label{ecc}\eeq

Various features of the paths and trajectories of the MS particle are graphically shown in Section 6.
\section{Trajectories of charged MS particles in external electric and magnetic field}

If we consider the electric field is along the x-axis and the magnetic field is along the z-axis, then the set of equations are
\beqa
\dot{p}_1 &=& eB\left(1+\frac{m}{\kappa}\right)v_2+eE\left(1+\frac{3m}{\kappa}\right)+\frac{e E m}{\kappa}v_1^2 \label{p1dot}\\ \dot{p}_2 &=& -eB\left(1+\frac{m}{\kappa}\right)v_1+\frac{e E m}{\kappa}v_1v_2 \label{p2dot}\\
\dot{p}_3 &=& \frac{e E m}{\kappa}v_1v_3\label{p3dot}.
\eeqa

Replacing $p$ by $m\dot{x}$ the above set of equations becomes
\beqa
m\ddot{x}_1 &=& eB\left(1+\frac{m}{\kappa}\right)\dot{x}_2+eE\left(1+\frac{3m}{\kappa}\right)+\frac{e E m}{\kappa}\dot{x}_1^2 \label{x1ddot}\\
m\ddot{x}_2 &=& -eB\left(1+\frac{m}{\kappa}\right)\dot{x}_1+\frac{e E m}{\kappa}\dot{x}_1\dot{x}_2 \label{x2ddot}\\
m\ddot{x}_3 &=& \frac{e E m}{\kappa}\dot{x}_1\dot{x}_3. \label{x3ddot}
\eeqa

The corresponding linear set of equation are
\beqa 
m\ddot{x}_1 &=& e B\left(1+\frac{m}{\kappa}\right)\dot{x}_2+e E\left(1+\frac{3m}{\kappa}\right) \label{x1linear}\\
m\ddot{x}_2 &=& -e B\left(1+\frac{m}{\kappa}\right)\dot{x}_1 \label{x2linear} \\
m\ddot{x}_3 &=& 0.\label{x3linear}
\eeqa

The solutions of these sets of equations are
\beqa
x_1 &=& \frac{c_1}{a_1\omega} Sin(a_1\omega t)-\frac{c_2}{a_1\omega} Cos(a_1\omega t)+c_3 \label{x1linear}\\
x_2 &=& \frac{c_1}{a_1\omega} Cos(a_1\omega t)+\frac{c_2}{a_1\omega} Sin(a_1\omega t)-\frac{Ea_2}{Ba_1}t+c_4 \label{x2linear}\\ x_3 &=& b_1+b_2 t, \label{x3linear}
\eeqa
where $\omega=\frac{eB}{m}$,$a_1=1+\frac{m}{\kappa}$ and $a_2=1+\frac{3m}{\kappa}$. To solve non-linear set of equations let us take the approximation
\beq
x_1=x_1^0+\frac{m}{\kappa}x_1^1~,~~x_2=x_2^0+\frac{m}{\kappa}x_2^1~,~~x_3=x_3^0+\frac{m}{\kappa}x_3^1 \label{x-approx} \eeq
where $x_i^0$ are solutions of linear set of equations. Setting this set of equations into the above equations (\ref{x1ddot}-\ref{x3ddot}) we get the differential equation for $x_i^1$'s as
\beqa 
\ddot{x}_1^1 &=& \omega a_1\dot{x}_2^1+\frac{E\omega}{B}(\dot{x}_1^0)^2 \label{x11ddot}\\
\ddot{x}_2^1 &=& -\omega a_1\dot{x}_1^1+\frac{E\omega}{B}\dot{x}_1^0\dot{x}_2^0 \label{x21ddot}\\
\ddot{x}_3^1 &=& \frac{E\omega}{B}\dot{x}_1^0\dot{x}_3^0 \label{x31ddot}
\eeqa
The solutions of this set of equations are
\beqa
x_1^1 &=& \frac{d_1}{a_1\omega}Sin(\theta)-\frac{d_2}{a_1\omega} Cos(\theta)-\frac{E(c_1^2-c_2^2)}{4B\omega a_1^2} Cos(2\theta)-\frac{E c_1c_2}{2B\omega a_1^2}Sin(2\theta)\nonumber\\
&&+\frac{3E^2 a_2}{4B^2\omega a_1^3}(c_2 Cos(\theta)-c_1 Sin(\theta))+\frac{E^2a_2
t}{2B^2 a_1^2}(c_1 Cos(\theta)+c_2 Sin(\theta))+d_3 \label{x11sol}\\
x_2^1 &=& \frac{d_1}{a_1\omega} Cos(\theta)+\frac{d_2}{a_1\omega} Sin(\theta)-\frac{E(c_2^2-c_1^2)}{4B\omega a_1^2}
Sin(2\theta)-\frac{E c_1c_2}{2B\omega a_1^2} Cos(2\theta)\nonumber\\
&&-\frac{E^2 a_2}{4B^2\omega a_1^3}(c_1 C o s(\theta)+c_2 Sin(\theta))-\frac{E^2a_2 t}{2B^2 a_1^2}
(c_1 Sin(\theta)-c_2 Cos(\theta))-\frac{E(c_1^2+c_2^2)}{2B a_1}t \label{x21sol} \\
x_3^1 &=& -\frac{E b_2}{B\omega a_1^2}(c_1 C o s(\theta)+c_2 Sin(\theta))+b_3 t+b_4 \label{x31sol}
\eeqa
The initial conditions are
\beqa
& x_1^0(0)=0~,~~x_2^0(0)=0~,~~x_3^0(0)=0~,~~x_1^1(0)=0~,~~x_2^1(0)=0~,~~x_3^1(0)=0~, &\nonumber\\
&\dot{x}_1^0(0)=v_1~,~~\dot{x}_2^0(0)=v_2~,~~\dot{x}_3^0(0)=v_3~,~~\dot{x}_1^1(0)=0~,~~\dot{x}_2^1(0)=0~,~~
\dot{x}_3^1(0)=0.& \label{inicon}
\eeqa
Using these initial conditions the values of the arbitrary constants are
\beqa & c_1=v_1~,~~c_2=v_2+\frac{E a_2}{B a_1}~,~~c_3=\frac{1}{\omega
a_1}(v_2+\frac{E a_2}{B a_1})~,~~c_4=-v_1~,
b_1=0~,~~b_2=v_3~,~~b_3=\frac{E v_3}{B a_1}(v_2+\frac{E a_2}{B
a_1})~,&\nonumber\\&
b_4=\frac{E v_1v_3}{B\omega a_1^2}~,~~d_1=\frac{E v_1v_2}{B a_1}+\frac{5E^2 a_2
v_1}{4B^2 a_1^2}~,~~d_2=\frac{E v_2^2}{B a_1}+\frac{7E^2 a_2 v_2}{4 B^2
a_1^2}+\frac{3E^3 a_2^2}{4B^3 a_1^3}~,&\nonumber\\&
d_3=\frac{3E v_2^2}{4B \omega a_1^2}+\frac{E v_1^2}{4B\omega a_1^2}+\frac{E^2
a_2 v_2}{2B^2\omega a_1^3}-\frac{E^3 a_2^2}{4B^3\omega a_1^4}~,~~d_4=-\frac{E
v_1 v_2}{2B\omega a_1^2}-\frac{E^2 v_1 a_2}{2B^2\omega a_1^3}~,~~&
\label{paravalue} \eeqa
Using (\ref{paravalue}) the solution of the set of equation(\ref{x1ddot}-\ref{x3ddot}) are
\beqa
x_1&=&\left[\frac{v_1}{\omega}+\frac{m}{\kappa}\left\{-\frac{v_1}{\omega}+\frac{E
v_1 v_2}{B\omega}+\frac{E^2 v_1}{2 B^2\omega}\right\}\right]Sin(\theta)
+\left[\left\{-\frac{v_2}{\omega}-\frac{E}{B\omega}\right\}+\frac{m}{\kappa}
\left\{\frac{v_2}{\omega}-\frac{E}{B\omega}-\frac{E v_2^2}{B\omega}-\frac{E^2
v_2}{B^2\omega}\right\}\right]\times\nonumber\\&& Cos(\theta)-\frac{m}{\kappa}\frac{E}{4B\omega}\left\{v_1^2-\left(v_2+\frac{E}{B}\right)^2\right\}Cos(2\theta)-
\frac{m}{\kappa}\frac{E v_1}{2B\omega}\left(v_2+\frac{E}{B}\right)Sin(2\theta)\nonumber\\&&
+\frac{m}{\kappa}\frac{E^2 v_1}{2B^2}t C o s(\theta)
+\frac{m}{\kappa}\frac{E^2}{2B^2}\left(v_2+\frac{E}{B}\right)t Sin(\theta)+\frac{v_2}{\omega}+\frac{E}{B\omega}\nonumber\\&&
+\frac{m}{\kappa}\left[-\frac{v_2}{\omega}+\frac{E}{B \omega}+\frac{3E
v_2^2}{4B \omega}+\frac{E v_1^2}{4B \omega}+\frac{E^2v_2}{2B^2\omega}-\frac{E^3}{4B^3\omega}\right] \label{x1}\\
x_2&=&\left[\frac{v_1}{\omega}+\frac{m}{\kappa}\left\{-\frac{v_1}{\omega}+\frac{Ev_1 v_2}{B\omega}+\frac{E^2 v_1}{B^2\omega}\right\}\right] Cos(\theta)+\left[\left\{\frac{v_2}{\omega}+\frac{E}{B\omega}\right\}+\frac{m}{\kappa}
\left\{-\frac{v_2}{\omega}+\frac{E}{B\omega}+\frac{E v_2^2}{B\omega}+\frac{3E^2v_2}{2B^2\omega}
\right.\right.\nonumber\\&& \left.\left.
+\frac{E^3}{2B^3\omega}\right\}\right]Sin(\theta)+\frac{m}{\kappa}\frac{E}{4B\omega}\left\{v_1^2-\left(v_2+\frac{E}{B}
\right)^2\right\}Sin(2\theta)-\frac{m}{\kappa}\frac{E v_1}{2B\omega}\left(v_2+\frac{E}{B}\right) Cos(2\theta)\nonumber\\&&
-\frac{m}{\kappa}\frac{E^2 v_1}{2B^2}t Sin(\theta)+\frac{m}{\kappa}\frac{E^2}{2B^2}\left(v_2+\frac{E}{B}\right)t C o s(\theta)-\left[\frac{E}{B}+\frac{m}{\kappa}\left\{\frac{2E}{B}+\frac{E}{2B}
\left(v_1^2+\left(v_2+\frac{E}{B}\right)^2\right)\right\}\right]t\nonumber\\&&
-\left[\frac{v_1}{\omega}+\frac{m}{\kappa}\left\{-\frac{v_1}{\omega}+\frac{E v_1v_2}{2B \omega}+\frac{E^2 v_1}{2B \omega}\right\}\right] \label{x2sol}\\ 
x_3&=&\left[v_3+\frac{m}{\kappa}\left\{\frac{E v_1 v_3}{B}+\frac{E^2v_3}{B^2}\right\}\right]t+\frac{m}{\kappa}\frac{E v_1
v_3}{B\omega}(1-Cos(\theta))-\frac{m}{\kappa}\frac{E v_3}{B\omega}\left(v_2+\frac{E}{B}\right)Sin(\theta)) \label{x3sol}
\eeqa
where $\theta=a_1\omega t$. If the velocity is along the z-axis then $v_1=v_2=0$ and $v_3=v$. Then we have
\beqa
x_1&=&\frac{E}{B\omega}\left[(1-C o s(\theta))-\frac{m}{\kappa} Cos(\theta)+\frac{m}{\kappa}\frac{E^2}{4B^2} Cos(2\theta)
+\frac{m}{\kappa}\frac{E^2}{2B^2}\omega t Sin(\theta)+\frac{m}{\kappa}\left(1-\frac{E^2}{4B^2}\right)\right] \label{x1}\\
x_2&=&-\frac{E}{B\omega}\left[(\omega t-Sin(\theta))-\frac{m}{\kappa}\left(1+\frac{E^2}{2B^2} \right)Sin(\theta)+\frac{m}{\kappa}\frac{E^2}{4B^2}Sin(2\theta)\right.\nonumber\\&& \left.
+\frac{m}{\kappa}\left(2+\frac{E^2}{2B^2}(1-Cos(\theta))\right)\omega t\right] \label{x2sol}\\
x_3&=&v_3\left[1+\frac{m}{\kappa}\frac{E^2}{B^2}\right]t-\frac{m}{\kappa}\frac{E^2v_3}{B^2\omega}Sin(\theta) \label{x3sol}
\eeqa
where $\theta=a_1\omega t$.
\section{Lienard-Wiechert potential for MS particle}
The full interacting action for electrodynamics of the conventional particle is
\beqa A&=&\int [-(d\tau ){\sqrt{(\dot X^\mu (\tau )\dot{X_\mu}(\tau )}}]-\int d^4y
~e~\delta (y-X(\tau ))\dot{X_\mu}A^\mu (y) -\frac{1}{4}\int d^4y~F^{\mu\nu
}(y)F_{\mu\nu }(y)\nonumber\\
&=&\int [-(d\tau ){\sqrt{(\dot X^\mu (\tau ) \dot{X_\mu}(\tau )}}]-\int (d\tau )
~e~\dot{X_\mu(\tau
)}A^\mu (X(\tau ))-\frac{1}{4}\int d^4y~F^{\mu\nu }(y)F_{\mu\nu }(y).
\eeqa

Now we notice that  the point particle and interaction sector depend on $X_\mu $ which are the dynamical variables. Indeed, 
when we consider the MS particle, $X_\mu$  need to be replaced by the physical $x_\mu$ variables that obey the NC algebra. On the other hand, the argument $y_\mu $ in the pure Maxwell sector are just  parameters and not dynamical variables which obviously are given by $A_\mu (y)$. Hence in the free Maxwell theory and resulting Maxwell equations there is no change
due to the MS particle. The only change that will appear is in the interacting system via the source term in the Maxwell equation
\beq
\partial^\mu F_{\mu\nu}=j_\nu (x),~ j_\nu =e v_\nu.
\label{m1}
\eeq

Now we can consider the Lienert-Wiechert potential for a moving charged MS particle. For the normal free charged particle this retarded potential is obtained from the inhomogeneous equation
\beq
\partial ^\nu\partial_\nu A^\mu =j^\mu ,
\label{2}
\eeq
in Lorentz gauge $\partial_\mu A^\mu =0$. This leads to the general solution
\beq
A^\mu (x)=A_0^\mu(x)+\int dx' G(x-x')j^\mu (x'),
\label{3}
\eeq
where $A_0^\mu(x)$ satisfies the homogeneous equation  $\partial^2 A_0^\mu(x)=0$ (see for example \cite{bar}). Using the retarded form of the Green's Function $G(x-x')$ one finds the potential for a current $j^\mu $,
\beq
A^\mu (x)=A_0^\mu(x)+\frac{1}{4\pi}\int dx' \frac{\delta (x^0-x^{'0}-\mid \vec
x-\vec x' \mid )}{\mid \vec x-\vec x' \mid} j^\mu (x').
\label{4}
\eeq

The form of current used by us is
\beq
j^\mu (x) = e\int d^4y \delta (y-x(\tau ))v_\mu (y).
\label{5}
\eeq

Hence, recalling the argument presented above, for the retarded potential for the MS particle we need only introduce the form of MS current which in turn means we should use the form of velocity appropriate for the MS particle.

We return to our equations of motion valid to $O(1/\kappa )$:
\beqa
\frac{dx^\alpha}{d\tau}&\sim& -\left(1+3\frac{(\eta p)}{\kappa}\right)\frac{p^\alpha
}{m_0}+\frac{e}{m_0\kappa}(\eta F p)x^\alpha,\nonumber\\
\frac{dp^\alpha}{d\tau}&\sim& \frac{e}{m_0}\left(1+\frac{(\eta p)}{\kappa}\right)F^{\alpha\beta }p_\beta
+\frac{e}{m_0\kappa}((\eta F p)p^\alpha +2m_0^2F^{\alpha\beta }\eta_\beta).
\label{6}
\eeqa

Again, to $O(e)$ and $O(1/\kappa )$, the above equations can be inverted to read
\beq
\frac{p^\alpha}{m_0}\sim -\left[\left(1-3\frac{(\eta
p)}{\kappa}\right)\frac{dx^\alpha}{d\tau}+\frac{e}{\kappa}\left(\eta F
\frac{dx}{d\tau}\right)x^\alpha\right].
\label{7}
\eeq

This motivates us to replace the velocity $(dx^\alpha )/(d\tau)$ in the charge
current by the MS velocity
\beq
\frac{dx^\alpha}{d\tau} \rightarrow \left[\left(1-3\frac{(\eta p)}{\kappa}\right)\frac{dx^\alpha}{d\tau}
+\frac{e}{\kappa}\left(\eta F\frac{dx}{d\tau}\right)x^\alpha\right].\label{8}
\eeq

First of all we consider the case with no external field, $F_{\mu\nu}=0$ so that the replacement is simply
\beq
\frac{dx^\alpha}{d\tau} \rightarrow
\left(1-3\frac{W}{\kappa}\right)\frac{dx^\alpha}{d\tau},\label{9}
\eeq
where the energy $W$ is computed from the MS dispersion relation (\ref{m}),
\beq
W=\left[-\frac{m_0^2}{\kappa}+\sqrt{\left(\frac{m_0^2}{\kappa}\right)^2+\left(1-\frac{m_0^2}{\kappa
^2}\right)(\vec p^2+m_0^2)}\right]/\left(1-\frac{m_0^2}{\kappa^2}\right).\label{10}
\eeq

To $O(\kappa )$ this simplifies to
\beq W\sim {\sqrt{(\vec p^2+m_0^2)}}-\frac{m_0^2}{\kappa}\label{11}
\eeq
with the rest energy being $ W_0 \sim \left(1-\frac{m_0}{\kappa }\right)m_0$. Hence, for a MS particle at rest, the factor $\left(1-3\frac{W_0}{\kappa}\right)\sim \left(1-3\frac{m_0}{\kappa}\right)$  modifies the Coulomb  potential,
\beq A_0=\frac{e}{4\pi R}\left(1-3\frac{m_0}{\kappa}\right);~~A_i=0. \label{12}
\eeq

Defining $\vec R=\vec x -\vec y(\tau _0),~\vec n=\vec R/R, \vec v=d\vec x /dt$, we find the situation is much more involved for a moving MS particle because now the factor $\left(1-3\frac{W}{\kappa}\right)$ is no longer a constant but depends on velocity,
\beq A_0=\frac{e}{4\pi R(1-\vec n.\vec v)}\left(1-3\frac{W}{\kappa}\right);~\vec A=\frac{e\vec v}{4\pi R(1-\vec n.\vec v)}\left(1-3\frac{W}{\kappa}\right). \label{13}
\eeq

However, probably more studies are needed to interpret our results for the MS particle induced retarded potential in the presence of external field $F_{\mu\nu}$ due to the presence of the $x^\alpha$-dependent term in (\ref{8}) in the expression for velocity. We will not pursue this analysis any further in the present publication.
\section{Discussion, conclusion and future outlook}

We will mainly study the results obtained in Section 4 regarding the paths and trajectories of MS particles in the external electromagnetic field. Let us demonstrate pictorially the $\kappa$-effects by the way of diagrams. For the sake of comparison we have always provided a diagram for $1/\kappa =0$ depicting the conventional particle. In the  Figure 1 we plot the particle energy as a function of $\tau$. To our approximation of $O(1/\kappa)$ the diagrams are not qualitatively different from the $1/\kappa=0$ (conventional particle) case but the energy increases more rapidly as $1/\kappa $ increases. Figure 2 shows the plot of $x_1$  vs. $\tau$ that is the coordinate in the direction of the external field. Once again the curves shows hyperbolic nature like conventional particle where $\kappa$ deformed curves becomes more steepy as $1/\kappa$ increases. However in Figure 3, where we plot $x_2$ vs. $\tau$, there appear qualitatively different paths for the MS particle in relation to the conventional particle. The latter is a simple straight line as there is no acceleration along $x_2$ in them conventional particle case but for MS particle  the hyperbolic nature dominates for later $\tau$ due to the $\kappa$-dependent acceleration term in $x_2$.

In the second set of figures (Figures 4,5 and 6) we show the trajectories in the $x_1-x_2$ plane for different choices of external parameters. In Figure 4, we plot the trajectories for different values of $\kappa$ for fixed external conditions: constant values of $\epsilon ~\sim$ electric field $E$ in the $x_1$-direction and constant $v_0~\sim$ initial energy of the particle. In Figure 5 we keep $v_0$ unchanged and consider the effect of varying $\epsilon$ on different $\kappa$-valued particles. Figure 6 shows the effect of different $v_0$ on different $\kappa$-valued particles where throughout $\epsilon$ is unaltered. Some common features of the trajectories are the following: quite expectedly the effects are more pronounced as $1/\kappa $ increases, as is apparent from Figure 4. In both figures 5 and 6, we have compared the behavior  of $1/\kappa =0.1$ particle with $1/\kappa =0$, the conventional particle. We notice that the $\kappa$-effects are more appreciable for larger values of the external parameters. This is also understandable since the external parameters are directly connected to the energy scales of the system and all $\kappa$-corrections involve the ratio of $(energy)/\kappa$. Lastly a general feature is that in all instances of fixed $\epsilon$ and $v_0$ the $\kappa$-particle trajectory bent more toward the $x_2$-axis with respect to the conventional $\kappa =0$ particle. This is because  $\kappa$-corrections produce a non-trivial acceleration in the $x_2$-direction even though the electric field is in $x_1$-direction, in contrast to the conventional particle for which the velocity in the $x_2$ direction remains constant. From the relation(\ref{ecc}) also it is clear that the eccentricity $\Sigma $ of the hyperbolic trajectory decreases as $1/\kappa$ increases.

Let us now comment on the connection between DSR principles and Quantum Gravity effects and a possible relevance of the present work in this perspective. From purely theoretical grounds, it appears that the existing framework of physical laws might require a significant modification in the regime of Quantum Gravity that is for Planck scale physics. Since it is expected that Planck scale provides the borderline of qualitatively different physics, (related to the nature of space-time  in particular), it is intuitively obvious that all inertial observers should agree on the same value of the Planck scale. Doubly Special Relativity (DSR) was developed by Amelino-Camelia \cite{am} keeping precisely this feature in mind. DSR is an extension of Einstein's Special Relativity. In Special Relativity there is one (inertial) observer independent dimensional parameter, the velocity of light but in DSR there are two observer independent dimensional parameters, the velocity of light  and an energy scale, conveniently chosen as the Planck energy (the role played by the parameter $\kappa$ in the present work). In this sense one might choose to interpret (in a purely phenomenological way) the $\kappa$-dependent modifications as Quantum Gravity effects but indeed, a rigorous justification of this identification should come from a first principle theory of Quantum Gravity, that is still lacking.

Finally let us summarize our work. We have studied the behavior of an extended model of charged particle, pertaining to the Magueijo-Smolin form of particle in the Doubly Special Relativity framework. The free particle resides in a phase space with $\kappa$-Minkowski form of Non-Commutative symplectic structure. As we have argued in the paper, studies of the above free particle model carried out so far in the literature is indeed interesting and important but the free particle cannot by itself provide a true picture. It is imperative to study the particle in the presence of non-trivial interactions as the interacting model can show the novelties as well as possible drawbacks (if there are any) of the new exotic particle models. In the present paper we have studied in detail the charged Magueijo-Smolin  particle interacting with electromagnetic field, both external as well as dynamical, in a perturbative framework. As we have demonstrated there are non-trivial $\kappa$-corrections in the particle trajectories. We have briefly discussed the dynamical gauge fields as well and how the Lienard-Wiechert type of potentials get modified due to the $\kappa$-corrected source term. Deeper analysis is required in the latter area.

For future work we intend to take up dynamics of the gauge field more thoroughly in the $\kappa$-Minkowski framework, especially in $2+1$-dimensions which is an area of recent interest \cite{new}. It is interesting to note that $2+1$-dimensional space-time allows more general form gauge kinetic terms other than just the Maxwell term, such as the Chern-Simons term or a combination of both the Maxwell and Chern-Simons term (that is the topologically massive gauge theory \cite{deser}). We will report on the MS particle dynamics in these forms of gauge kinetic terms in future publications.

\begin{figure}[htb]
{\centerline{\includegraphics[width=7cm, height=4cm] {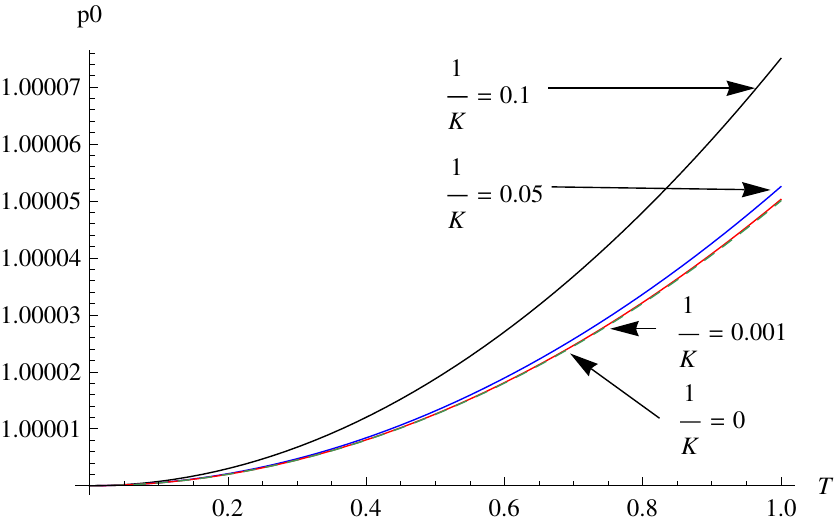}}}
\caption{{\it{Plot of $p_0$ vs. $\tau$ for fixed $\epsilon=0.01,m=1$ and
different $\kappa$ with  $\frac{1}{\kappa}=0$ for broken line.}}} \label{fig1}
\end{figure}
\begin{figure}
{\centerline{\includegraphics[width=7cm, height=4cm] {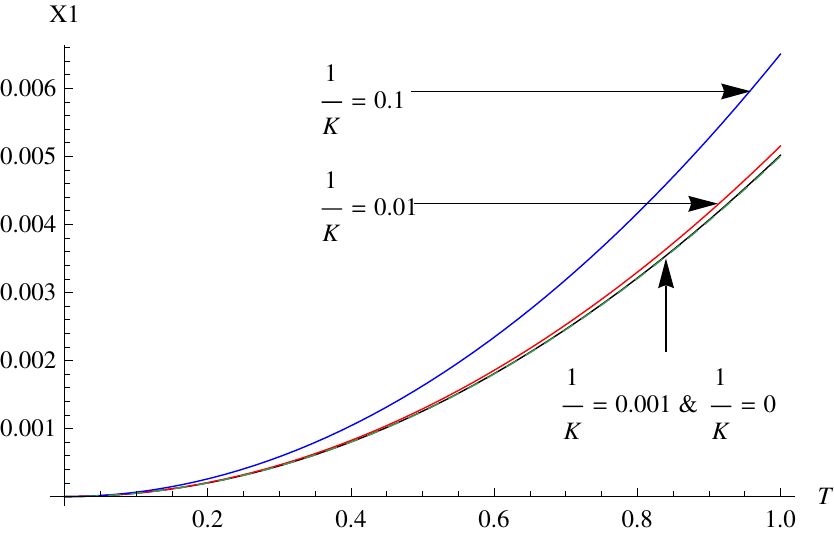}}}
\caption{{\it{Plot of $x_1$ vs. $\tau$ for fixed $\epsilon=0.01,m=1$ and
different $\kappa$ with  $\frac{1}{\kappa}=0$ for broken line.}}} \label{fig2}
\end{figure}
\begin{figure}
{\centerline{\includegraphics[width=7cm, height=4cm] {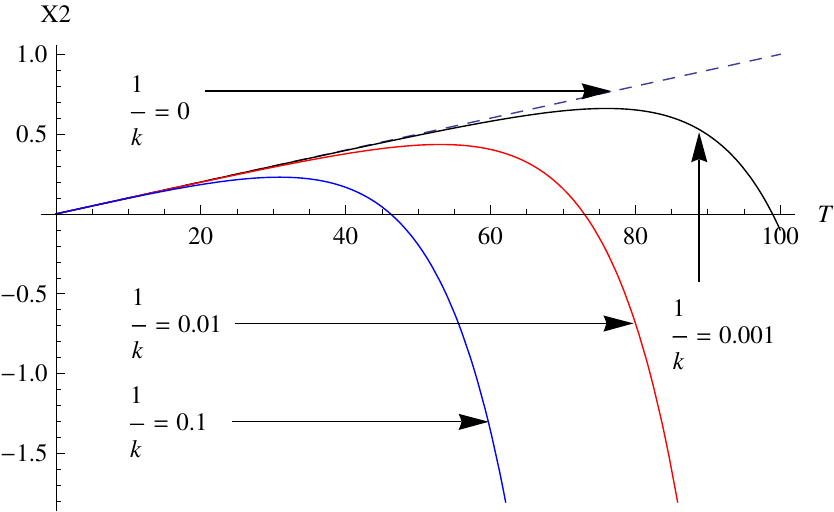}}}
\caption{{\it{Plot of $x_2$ vs. $\tau$ for fixed $\epsilon=0.1,m=1$ and
different $\kappa$ with  $\frac{1}{\kappa}=0$ for broken line.}}} \label{fig3}
\end{figure}
\begin{figure}
{\centerline{\includegraphics[width=10cm, height=3cm] {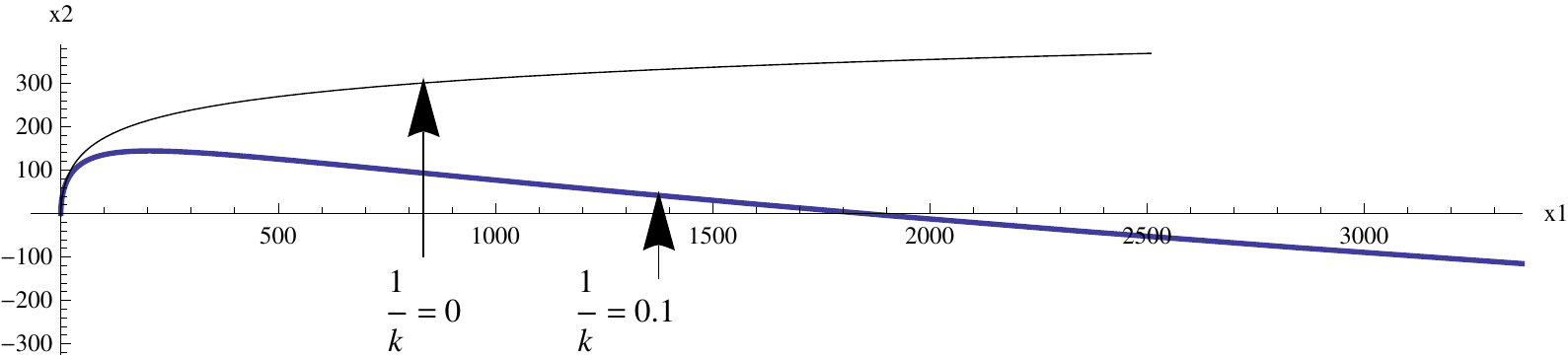}}}
\caption{{\it{Trajectory of the particle for
initial velocity $v_0=0.5$ and $\epsilon=0.08$}}} \label{fig4}
\end{figure}
\begin{figure}
{\centerline{\includegraphics[width=7cm, height=4cm] {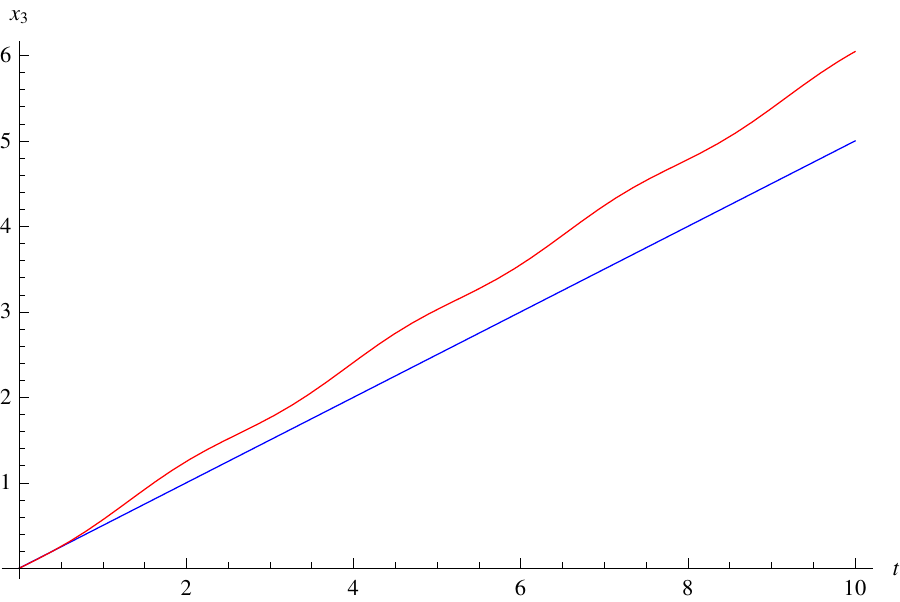}}}
\caption{{\it{Plot for $x_3$ vs $t$ with initial velocity $v_3=0.5$ and for fixed $\omega=2$ and $\frac{E}{B}=1$.
Blue and Red line are corresponding to $\frac{1}{\kappa}=0$ and $0.2$ respectively.}}} \label{fig5}
\end{figure}
\begin{figure}
{\centerline{\includegraphics[width=10cm, height=3cm] {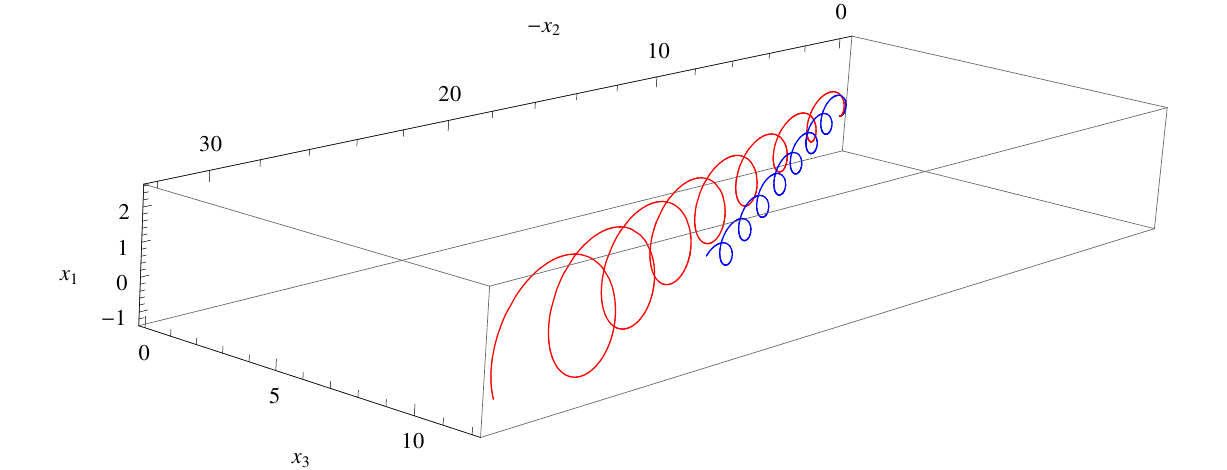}}}
\caption{{\it{Trajectory of the particle with initial velocity $v_3=0.5$ for fixed $\omega=2$ and $\frac{E}{B}=1$.
Blue and Red line are corresponding to $\frac{1}{\kappa}=0$ and $0.2$ respectively.}}} \label{fig6}
\end{figure}
\newpage
\section{Appendix}
We provide some  computational  details for the solutions of  the system of
non-linear ordinary differential equation (\ref{d2p0}-\ref{d2p3}),
\beq \frac{d^2
p_0}{d\tau^2}=\epsilon^2\left[p_0+\frac{1}{\kappa}
\left(3p_0^2+p_1^2+2m^2\right)\right],~
 \frac{d^2
p_1}{d\tau^2}=\epsilon^2\left[p_1+\frac{2}{\kappa}p_0p_1\right],\label{d2p1}\eeq
\beq \frac{d^2 p_2}{d\tau^2}=-\frac{\epsilon^2}{\kappa}p_0p_2,~
 \frac{d^2 p_3}{d\tau^2}=-\frac{\epsilon^2}{\kappa}p_0p_3\label{d2p3a}\eeq
 We  solve it in a perturbative framework in the first non-trivial order in
$1/\kappa $. Let us assume $p_{00},~p_{10},~p_{20},~p_{30}$ to be the
solution of $p_{0},~p_{1},~p_{2},~p_{3}$, for $\kappa\rightarrow\infty$, that is
the conventional case. Now
 we expand  $p_\mu$ about the conventional solutions $p_{\mu 0}$:
\beq
p_0=p_{00}+\frac{c}{\kappa}p_{01},~p_1=p_{10}+\frac{d}{\kappa}p_{11},~p_2=p_{20}
+\frac{g}{\kappa}p_{21},~p_3=p_{30}+\frac{h}{\kappa}p_{31}
\label{perturequ}\eeq
Here $c,d,g,h$ are constants to be evaluated later. If at $\tau=0$ the initial
velocity $v_0$ is along the $y$-axis then initial conditions for the above set of perturbation are
\beq p_0(0)=m~\Rightarrow~p_{00}(0)=m~,~p_{01}(0)=0~,$$$$
p_1(0)=0~\Rightarrow~p_{10}(0)=p_{11}(0)=0~,$$$$
p_2(0)=mv_0~\Rightarrow~p_{20}(0)=mv_0~,~p_{21}(0)=0~,$$$$
p_3(0)=0~\Rightarrow~p_{30}(0)=p_{31}(0)=0 \label{initialcon} \eeq
Using (\ref{perturequ}) the above set of differential equations becomes
\beq \frac{d^2 p_{00}}{d\tau^2}=\epsilon^2p_{00},~\frac{d^2
p_{10}}{d\tau^2}=\epsilon^2p_{10},~
\frac{d^2 p_{20}}{d\tau^2}=0 ,~\frac{d^2 p_{30}}{d\tau^2}=0,
\label{d2p00}\eeq
\beq c\frac{d^2 p_{01}}{d\tau^2}=\epsilon^2\left[c
p_{01}+3p_{00}^2+p_{10}^2+2m^2\right],\label{d2p01}\eeq
\beq d\frac{d^2 p_{11}}{d\tau^2}=\epsilon^2\left[d
p_{11}+2p_{00}+p_{10}\right],\label{d2p11}\eeq
\beq g\frac{d^2 p_{21}}{d\tau^2}=-\epsilon^2p_{00}p_{20},~
  h\frac{d^2 p_{31}}{d\tau^2}=-\epsilon^2p_{00}p_{30}.\label{d2p31}\eeq
The solution of (\ref{d2p00}) are respectively
\beq p_{00}=a_1e^{\epsilon\tau}+a_2e^{-\epsilon\tau},~p_{10}=b_1e^{\epsilon\tau}+b_2e^{-\epsilon\tau}\label{solp10}\eeq
\beq p_{20}=c_1+c_2\tau,~ p_{30}=d_1+d_2\tau\label{solp30}\eeq
Now using (\ref{solp10}) and (\ref{solp30}) the solutions of the differential equation (\ref{d2p01}),(\ref{d2p11})
and (\ref{d2p31}) are respectively
\beq
p_{01}=l_1e^{\epsilon\tau}+l_2e^{-\epsilon\tau}+\frac{3a_1^2+b_1^2}{3c}e^{
2\epsilon\tau}+\frac{3a_2^2+b_2^2}{3c}e^{-2\epsilon\tau}
-\frac{6a_1a_2+2b_1b_2+2m^2}{c}\label{soup01}\eeq
\beq
p_{11}=m_1e^{\epsilon\tau}+m_2e^{-\epsilon\tau}+\frac{2a_1b_1}{3D}e^{
2\epsilon\tau}+\frac{2a_2b_2}{3d}e^{-2\epsilon\tau}
-\frac{2(a_1b_2+b_1a_2)}{d}\label{soup11}\eeq
\beq p_{21}=-\frac{1}{g}\left[\left(a_1c_1-\frac{2a_1c_2}{\epsilon}+\tau
a_1c_2\right)e^{\epsilon\tau}+\left(a_2c_1+\frac{2a_2c_2}{\epsilon}+\tau
a_2c_2\right)e^{-\epsilon\tau}\right]+c'\tau+c''\label{solp21}\eeq
\beq
p_{31}=-\frac{1}{h}\left[\left(\tau-\frac{2}{\epsilon}\right)a_1d_2e^{
\epsilon\tau}+\left(\tau+\frac{2}{\epsilon}\right)a_2d_2e^{-\epsilon\tau}\right]
+c'''\tau+c''''\label{solp31}\eeq
Therefore
\beq
p_0=\left(a_1+\frac{cl_1}{\kappa}\right)e^{\epsilon\tau}+\left(a_2+\frac{cl_2}{
\kappa}\right)e^{-\epsilon\tau}+\frac{3a_1^2+b_1^2}{3\kappa}
e^{2\epsilon\tau}+\frac{3a_2^2+b_2^2}{3\kappa}e^{-2\epsilon\tau}-\frac{
6a_1a_2+2b_1b_2+2m^2}{c\kappa}\label{solp0gen}\eeq
\beq
p_1=\left(b_1+\frac{dm_1}{\kappa}\right)e^{\epsilon\tau}+\left(b_2+\frac{dm_2}{
\kappa}\right)e^{-\epsilon\tau}+\frac{2a_1b_1}{3\kappa}e^{2\epsilon\tau}
+\frac{2a_2b_2}{3\kappa}e^{-2\epsilon\tau}-\frac{2(a_1b_2+b_1a_2)}{\kappa}\label
{solp1gen}\eeq
\beq p_2=c_1+c_2\tau-\frac{1}{\kappa}\left[\left(a_1c_1+\tau
a_1c_2-\frac{2a_1c_2}{\epsilon}\right)e^{\epsilon\tau}+\left(a_2c_1+\tau
a_2c_2+\frac{2a_2c_2}{\epsilon}\right)e^{-\epsilon\tau}\right]+\frac{c'g\tau}{
\kappa}+\frac{c''g}{\kappa}\label{solp2gen}\eeq
\beq
p_3=d_2\tau-\frac{1}{\kappa}\left[\left(\tau-\frac{2}{\epsilon}\right)a_1d_2e^{
\epsilon\tau}+\left(\tau+\frac{2}{\epsilon}\right)a_2d_2e^{-\epsilon\tau}
\right]+\frac{c'''h\tau}{\kappa}+\frac{c''''h}{\kappa}\label{solp3gen}\eeq
All the arbitrary constants introduced above are solved by checking consistency
of the set of solutions leading to
the result given below:
\beq
a_1=a_2=\frac{m}{2}~,~b_1=-b_2=\frac{m}{2}~,~cl_1=cl_2=\frac{m^2}{6}+m^2~,
~dm_1=-dm_2=\frac{m^2}{6}+m^2~,$$$$
~c_1=mv_0~,~c_2=0~,~c'=0~,~c''=\frac{c_1m}{g}~,~d_1=0~,~d_2=0~,~c'''=0~,
~c''''=0\label{parametervalue}\eeq

\ed